\documentclass[longbibliography,prl,aps,twocolumn,showpacs,floatfix]{revtex4-2}
\usepackage{amsfonts}
\usepackage{amssymb}
\usepackage{hyperref}
\usepackage{graphicx}
\usepackage{dcolumn}
\usepackage{bm,amsmath,verbatim}
\usepackage{mathrsfs}
\usepackage{color}
\usepackage{lineno}
\usepackage{dsfont}
\usepackage{amsmath,amssymb,amsthm}

\usepackage[T1]{fontenc}
\usepackage[latin9]{inputenc}
\usepackage{booktabs}

\hypersetup{colorlinks,
	linkcolor=blue,          citecolor=blue,        filecolor=blue,      urlcolor=blue           }

\def\be{\begin{equation}}
	\def\ee{\end{equation}}
\def\bea{\begin{eqnarray}}
	\def\eea{\end{eqnarray}}
\def\bse{\begin{subequations}}
	\def\ese{\end{subequations}}

\def\be{\begin{eqnarray}}
	\def\ee{\end{eqnarray}}

\begin{document}
	\title{Nonlinearity-induced Fractional Thouless Pumping of Solitons}
	\author{Yu-Liang Tao$^{1}$}
	\author{Yongping Zhang$^{2}$}
	\email{yongping11@t.shu.edu.cn}
	\author{Yong Xu$^{1,3}$}
	\email{yongxuphy@tsinghua.edu.cn}
	\affiliation{$^{1}$Center for Quantum Information, IIIS, Tsinghua University, Beijing 100084, People's Republic of China}
	\affiliation{$^{2}$Department of Physics, Shanghai University, Shanghai 200444, People's Republic of China}
	\affiliation{$^{3}$Hefei National Laboratory, Hefei 230088, People's Republic of China}
	
	\begin{abstract}
		Recent studies have shown that a soliton can be {\it fractionally} transported by slowly varying a system 
		parameter over one period in a nonlinear system. This phenomenon is attributed to the nontrivial topology of 
		the corresponding energy bands of a linear Hamiltonian. 
		Here we find the occurrence of fractional Thouless pumping of solitons in a nonlinear off-diagonal 
		Aubry-Andr\'{e}-Harper model. Surprisingly, this happens despite the fact that all the energy bands of 
		the linear Hamiltonian are topologically trivial, indicating that nonlinearity can induce fractional 
		Thouless pumping of solitons. Specifically, our results show that a soliton can be pumped across one unit cell 
		over one, two, three or four pump periods, implying an average displacement of $1$, $1/2$, $1/3$ or $1/4$ unit cells
		per cycle, respectively.
		We attribute these behaviors to changes in on-site potentials induced by a soliton solution, leading to
		the nontrivial topology for the modified linear Hamiltonian. Given that our model relies solely on 
		varying nearest-neighbor hoppings, it is readily implementable on existing state-of-the-art photonic platforms.
	\end{abstract}
	
	\maketitle
	Nonlinearity is ubiquitous in various distinct disciplines, such
	as physics, biology, chemistry, economics and social sciences.
	In physics, for instance, 
	nonlinearity plays a crucial role in photonic systems~\cite{christodoulidesOL1988,eisenbergPRL1998,fleischerNat2003,christodoulidesNat2003,kivsharbook2003,ledererPR2008discrete,kevrekidisBook2009} 
	and Bose-Einstein condensates (BECs)~\cite{brazhnyiMPLB2004,morschRMP2006,dauxoisbook2006,BEC_book_2008,chinRMP2010,
		busch2000motion,liu2018nonlinear},
	leading to many fascinating phenomena, 
	including chaos~\cite{lorenz1963,may1976,gutzwiller1990} and solitons~\cite{kivsharbook2003,kevrekidisBook2009,dauxoisbook2006,BEC_book_2008}. 
	Here, a soliton refers to a wave packet whose shape remains unchanged while traveling
	due to the balance between dispersion and nonlinearity.
	Recently, nonlinear systems meet topology, resulting in significant
	discoveries, such as topological bulk~\cite{lumerPRL2013,mukherjeeSci2020} and edge solitons~\cite{ablowitzPRA2014,leykamPRL2016,mukherjeePRX2021,tao2020hinge}, and nonlinearity-induced 
	topological phase transitions~\cite{maczewskySci2020,soneNP2024}. 
	
	In particular, recent studies have demonstrated that in a nonlinear system, 
	a soliton can undergo quantized transport when a linear Hamiltonian is slowly 
	varied in a periodic manner~\cite{jurgensenNat2021,jurgensenPRL2022,fuPRL2022,fu2022twoD,mostaanNC2022,tuloupNJP2023,citro2023thouless,
		HuNJP2024,lyuPRR2024,szameit2024discrete}.
	This phenomenon is in stark contrast to the linear Thouless pumping in electronic systems~\cite{thoulessPRB1983,niu1984quantised},
	where an entire linear band is required to be occupied. Despite the difference, the quantized 
	transport is found to be dictated by the Chern number of a Bloch band of the linear 
	Hamiltonian~\cite{jurgensenPRL2022,mostaanNC2022,fuPRL2022}, from which the soliton bifurcates. 
	Additionally, it has been found that a soliton can be pumped across a unit cell
	over more than one cycles, a process referred to as fractional Thouless pumping of solitons~\cite{jurgensenNP2023}. 
	This pumping is now understood as the flow of the instantaneous maximally
	localized multi-band Wannier functions of the linear Hamiltonian~\cite{jurgensenNP2023}. 
	As a result, if the total 
	Chern number of the corresponding bands vanishes, a soliton cannot experience transport.
	Remarkably, a recent work has predicted the breakdown of the correspondence between 
	the displacement of a soliton and the Chern number for an {\it integer} nonlinear Thouless
	pump~\cite{tao2024nonlinearity}. However, it remains unclear whether fractional Thouless pumping is able to occur 
	for a soliton when the linear Hamiltonian is topologically trivial.   
	
	\begin{figure}
		\includegraphics[width=1\linewidth]{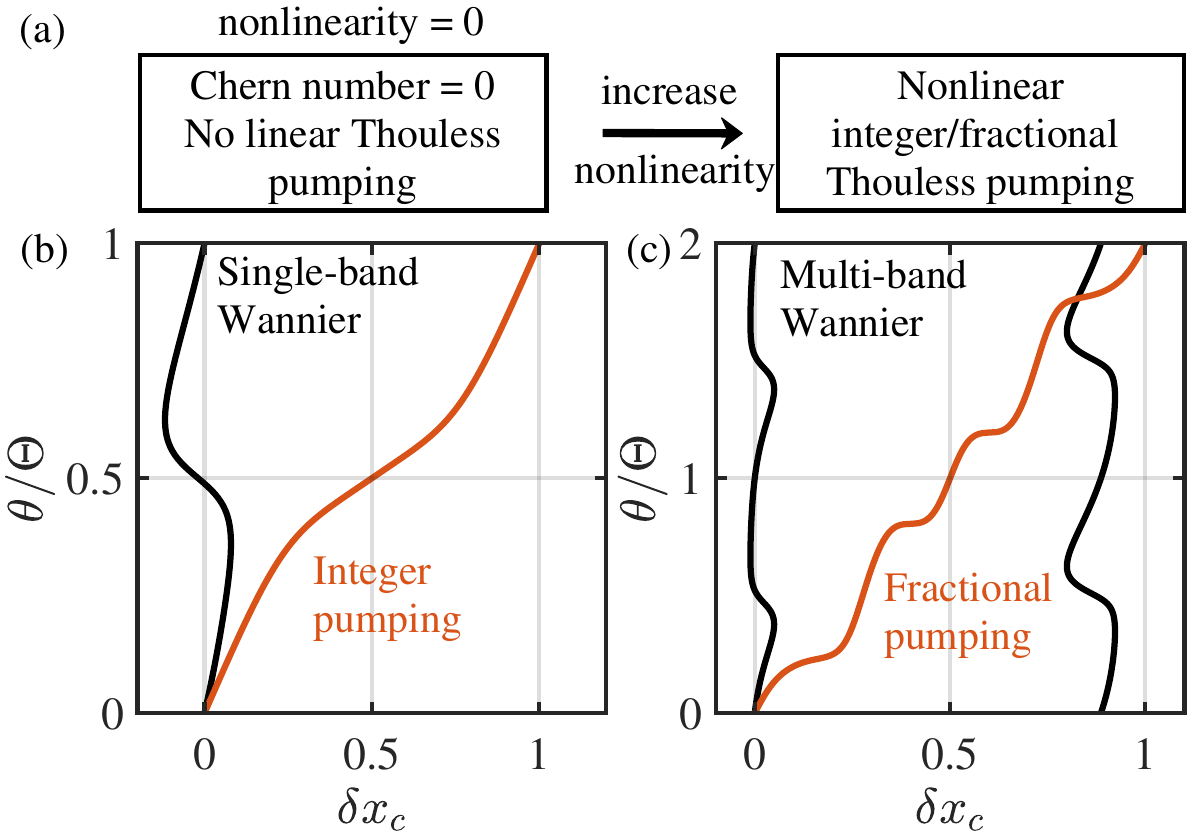}
		\caption{(a) Schematic illustration of nonlinearity-induced Thouless pumping. 
			In the absence of nonlinearity, linear Thouless pumping does not occur if the Chern number 
			of the relevant energy bands is zero.
			However, introducing nonlinearity allows a soliton to undergo integer or fractional Thouless 
			pumping.
			The trajectories of the relative center-of-mass positions of 
			the instantaneous solitons (red lines) 
			and the relevant instantaneous Wannier functions (black lines) for nonlinear (b) integer 
			and (c) fractional pumping. Despite the absence of flow in the Wannier functions, 
			nonlinear integer or fractional Thouless pumping can be observed.}
		\label{Fig_1}
	\end{figure}
	
	Here we study a nonlinear off-diagonal Aubry-Andr\'{e}-Harper (AAH) model, finding
	that a soliton exhibits integer quantized transport when a system parameter
	in a linear Hamiltonian is slowly tuned over a period, even if the linear Hamiltonian itself 
	is topologically trivial (see Fig.~\ref{Fig_1}). 
	Unlike the model in Ref.~\cite{tao2024nonlinearity}, in this case, nonlinear coefficients are not 
	required to change during a pump period, thereby significantly simplifying experimental implementation. 
	We further demonstrate that in topologically trivial AAH models,
	a soliton can be transported across one unit cell over {\it two}, {\it three} or {\it four} cycles,  
	indicating that nonlinearity is able to induce fractional Thouless pumping for 
	a trivial linear Hamiltonian.  
	We attribute these phenomena to modifications in on-site potentials induced by
	soliton solutions, which effectively transform the linear Hamiltonian into a topologically 
	nontrivial form. 
	As the proposed method only involves varying nearest-neighbor hoppings, it can be 
	readily implemented on current state-of-the-art photonic platforms.

\begin{figure}
		\includegraphics[width=1\linewidth]{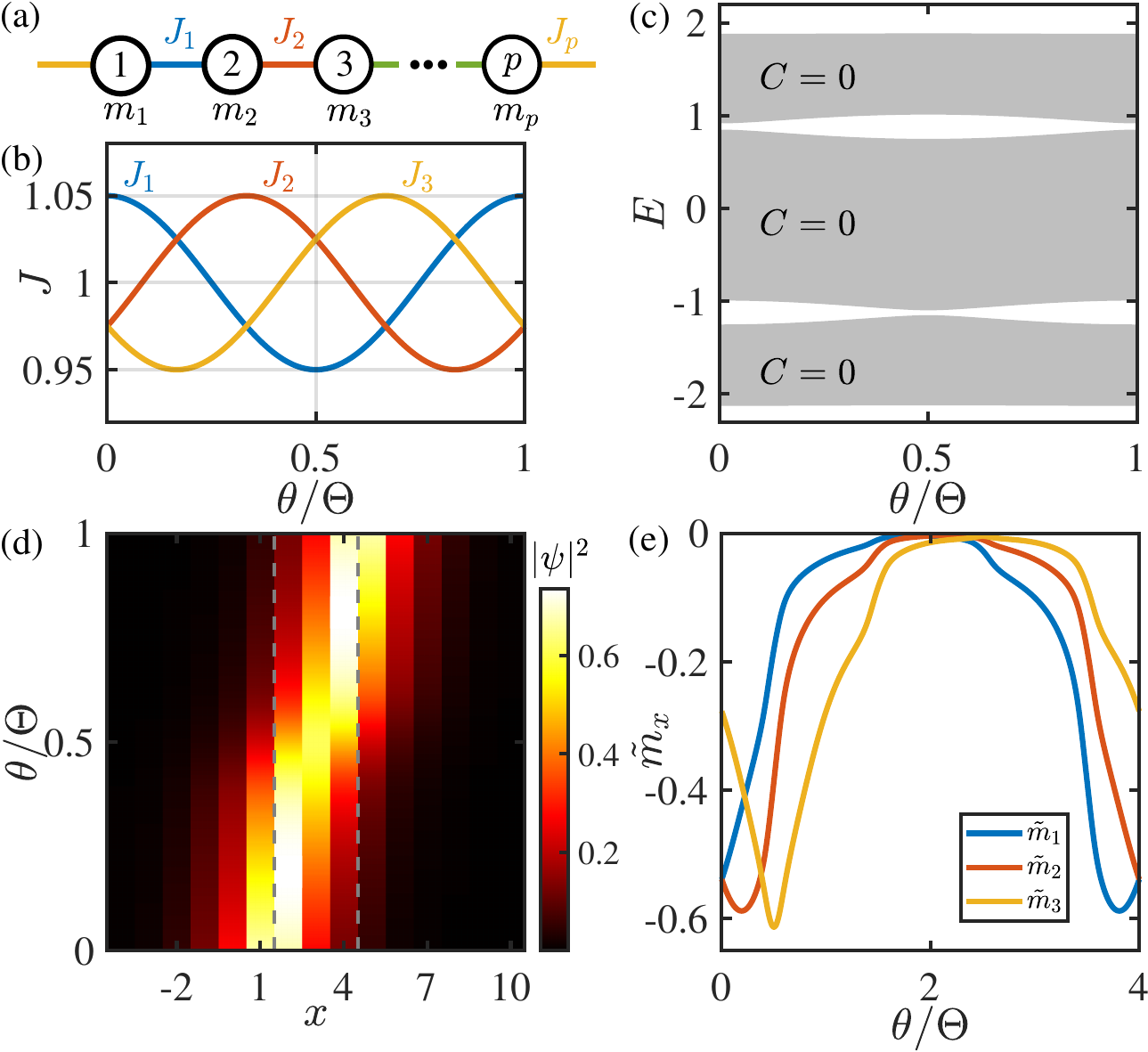}
		\caption{(a) Schematics of the off-diagonal AAH model, a 1D chain with $p$ sites per unit cell. 
			Besides the nearest-neighbor hoppings, the model contains on-site potentials $m_x$ that do not change over time.
			(b) The hopping strength as a function of $\theta$ during one period for the case with $p=3$ and $q=1$. 
			(c) Band structures of the off-diagonal AAH model in Eq.~(\ref{H_TB}) with $p=3$ and $q=1$ 
			with respect to the parameter $\theta$. There are three topologically trivial bands with zero Chern numbers. 
			(d) The evolution of the density distribution $|\psi_x|^2$ of instantaneous solitons bifurcating from the 
			lowest linear band as $\theta$ changes from $0$ to $\Theta$. The dashed lines highlight the center-of-mass 
			positions of the solitons at $\theta=0$ and $\theta=\Theta$. 
			(e) The on-site potentials $\tilde{m}_x(\theta)=g_x|\psi_x(\theta)|^2$ ($x=1,2,3$) induced by 
			soliton solutions versus $\theta$ over four periods, which are determined by the 
			instantaneous solitons in (d). Here, we set $N=2.3$, $J_a=0.05$, $\{m_x\}=\{-0.2,-0.2,0.04\}$, 
			and $\{g_x\}=\{-0.8,-0.8,-1\}$ with $x=1,2,3$.}
		\label{Fig_2}
	\end{figure}

	\emph{Model}.---To demonstrate nonlinearity-induced Thouless pumping of solitons, 
	we start by considering an off-diagonal AAH model~\cite{AAH_1,AAH_2,krausPRL2012,lang2012edge,liuPRB2015,keLPR2016}, 
	whose energy bands are topologically trivial.
	The system is a one-dimensional (1D) chain with nearest-neighbor hoppings and 
	on-site potentials $m_x$ [see Fig.~\ref{Fig_2}(a)] described by the following Hamiltonian 
	\begin{equation}
		\label{H_TB}
		H^{\textrm{lin}}_{x^\prime,x}=-J_{x^\prime}\delta_{x^\prime+1,x}-J_{x}\delta_{x^\prime-1,x}+m_x\delta_{x^\prime,x},
	\end{equation}
	where $x$ and $x^\prime$ denote lattice sites, and  
	$J_{x}=J_{[(x-1) \text{ mod } p]+1}$ 
	is the hopping strength between sites $x$ and $x+1$.
	Due to different values of $m_x$, there are $p$ sites in each unit cell.
	The nearest-neighbor hoppings are periodically modulated in $\theta$ with period $\Theta$, taking the following form
	\begin{equation}
		\label{J_x}
		J_{x}(\theta)=J_0+J_a \cos \left( {2\pi}\theta/{\Theta}-{2\pi q}(x-1)/{p}\right),
	\end{equation} 
	where $J_a$ is the oscillation amplitude of the modulated hoppings, 
	and $J_0$ is the hopping's central strength. For simplicity, we set $J_0=1$ as the unit of energy.
	Here, the positive integer $q$ (which is prime to $p$ and $q<p/2$) describes the relation 
	between adjacent nearest-neighbor hoppings. This model has $p$ energy bands characterized by a 
	momentum $k$. Besides this $k$, we can view $\theta^\prime=2\pi \theta/\Theta$ as the other momentum,
	so that the Chern number of a Bloch band with respect to $k$ and $\theta^\prime$ can be evaluated. 
	Previous studies show that this Chern number is essential for both integer and fractional Thouless pumping of
	solitons~\cite{jurgensenNat2021,jurgensenPRL2022,fuPRL2022,mostaanNC2022,jurgensenNP2023}.
	
	In the presence of on-site nonlinearity, the dynamics of the system is governed by the following dimensionless discrete 
	nonlinear Schr{\"o}dinger equation:
	\begin{equation}
		\label{NLSETB}
		i\frac{\partial}{\partial t} \psi_x=\sum_{x^\prime}H^{\textrm{lin}}_{x,x^\prime}(\theta)\psi_{x^\prime}+g_x|\psi_x|^2\psi_x,
	\end{equation} 
	where $\psi_x$ and $g_x$ are the value of a wavefunction and the nonlinear coefficient at site $x$, respectively.
	In photonic systems, $t$ denotes the propagation distance.
	The parameter $\theta$ is slowly varied over time.
	The norm of the wavefunction $N=\sum_{x}|\psi_x|^2$ represents the strength of nonlinearity,
	and it remains constant during time evolution.
	In addition, unlike the proposal in Ref.~\cite{tao2024nonlinearity}, $g_x$ stays constant as $t$ changes, thus rendering
	it more readily accessible for experiments. In fact, our proposal only requires the variation of hopping,
	while the on-site potentials remains unchanged throughout the time evolution process, (i.e., they are independent of $\theta$), 
	a condition that can be immediately realized using the state-of-art photonic technology~\cite{jurgensenNat2021,jurgensenNP2023}.

	\emph{Nonlinearity-induced integer Thouless pumping of solitons}.---To demonstrate this,
	we consider the case with $p=3$ and $q=1$. Without loss of generality, we set $J_a=0.05$ 
	and $\{m_x\}=\{-0.2,-0.2,0.04\}$ with $x=1,2,3$, 
	and modulate the hoppings by varying $\theta$ based on Eq.~(\ref{J_x}) [see Fig.~\ref{Fig_2}(b)]. This model
	has three bands, all of which are topologically trivial with the Chern number $C=0$ 
	[see Fig.~\ref{Fig_2}(c)]. To generate solitons, we introduce site-dependent focusing nonlinearity 
	with $\{g_x\}=\{-0.8,-0.8,-1\}$ ($x=1,2,3$). 
	We compute the stable instantaneous soliton solution of the instantaneous nonlinear Hamiltonian 
	at each $\theta$ using the Newton's method. These solutions emerge from the lowest linear band.
	Note that, in the adiabatic limit, where $\theta$ is varied very slowly, these solutions 
	represent the evolution of the soliton over time~\cite{kivsharRMP1989,bandPRA2002,bandPRA2002_2,liuPRL2003,wuPRL2005}.

	Remarkably, Fig.~\ref{Fig_2}(d) illustrates that a soliton is displaced by one unit cell (three sites) as 
	$\theta$ is modulated from $0$ to $\Theta$, 
	although the corresponding linear band is topologically trivial. 
	The displacement of the soliton is more clearly depicted by the relative movement 
	of the soliton's center of mass in units of the length of a unit cell 
	defined as
	$
	\delta x_c(\theta)=[x_c(\theta)-x_c(0)]/p,
	$ 
	with $x_c(\theta)=\sum_{x}x|\psi_x(\theta)|^2/N$ [see the red line in Fig.~\ref{Fig_1}(b)]. 
	It starkly contrasts with the motion of the Wannier center of the corresponding linear band, 
	which returns to its initial position after one period [see the black line in Fig.~\ref{Fig_1}(b)]. 
	The finding indicates that 
	nonlinearity can enable integer Thouless pumping of a soliton bifurcating from a linear trivial band, 
	significantly simplifying implementation as
	nonlinearity is not required to change during time evolution.
	
	\begin{figure}
		\includegraphics[width=1\linewidth]{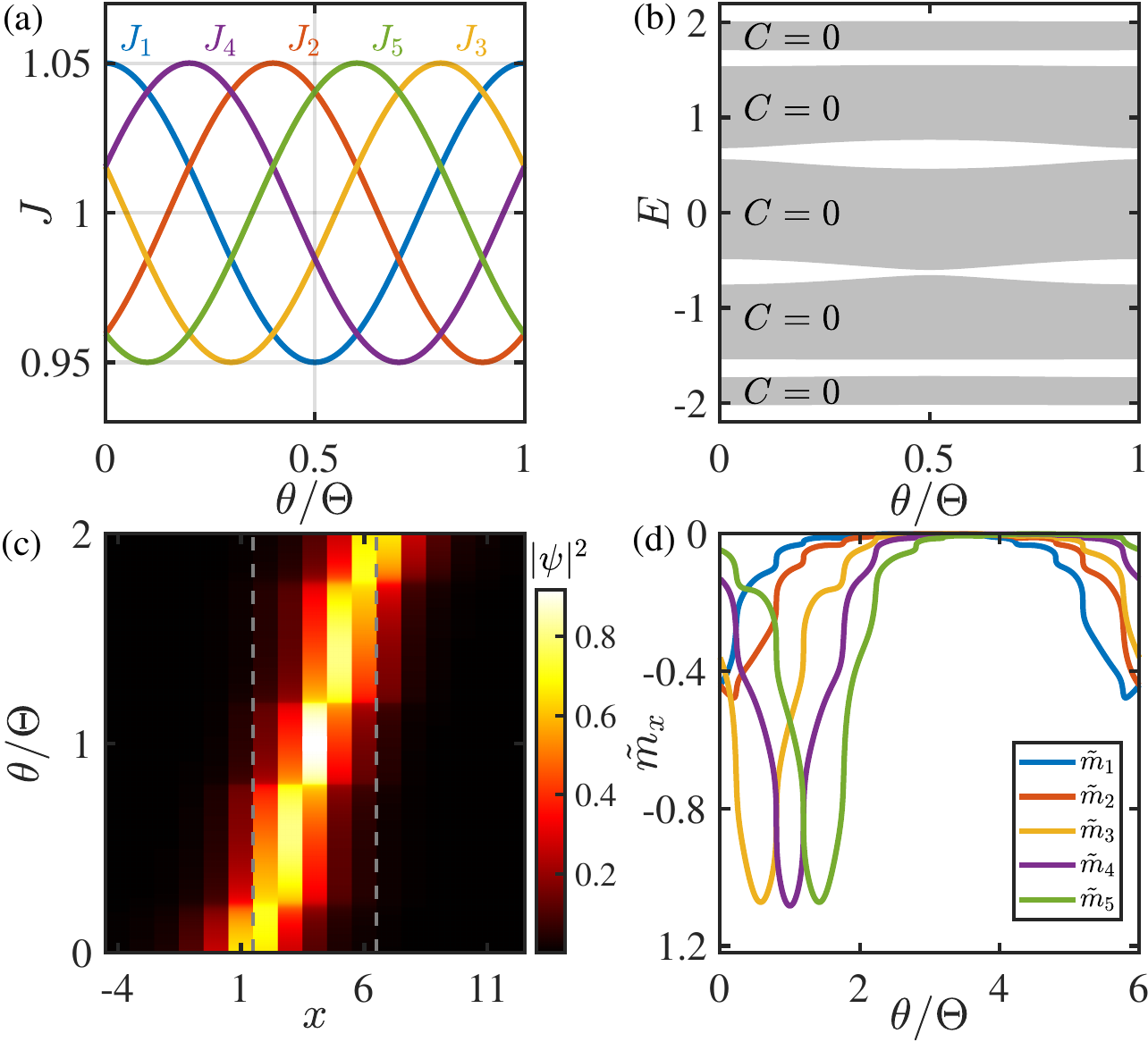}
		\caption{Nonlinearity-induced fractional Thouless pumping where a soliton 
			is pumped across one unit cell (five sites) after two periods for the case with $p=5$ and $q=2$.
			(a) The hopping strength versus $\theta$.
			(b) One-cycle band structures.
			(c) The evolution of the density distribution $|\psi_x|^2$ of instantaneous solitons bifurcating 
			from the lowest two linear bands over two periods. 
			The dashed lines mark the center-of-mass positions of the soliton at $\theta=0$ and $\theta=2\Theta$, respectively.
			(d) Nonlinear on-site potentials $\tilde{m}_x(\theta)$ versus $\theta$ over six periods, 
			which is determined by the instantaneous soliton solutions shown in (c). 
			Here, we take $N=2.2$, $J_a=0.05$, 
			$\{m_x\}=\{-0.2,-0.2,0.2,-0.02,0.2\}$, and $\{g_x\}=\{-0.67,-0.67,-1.33,-1.18,-1.33\}$.
		}
		\label{Fig_3}
	\end{figure}
	
	We now interpret the nonlinearity-induced pumping as arising from modifications of 
	on-site potentials by a soliton solution,
	resulting in a modified linear Hamiltonian that is topologically nontrivial. 
	Specifically, consider the case that a soliton is displaced by $n$ unit cells over $m$ cycles
	(e.g., $n=1$ and $m=1$).
	For an integer $l$ that is an integer multiple of $m$,
	as $\theta$ is slowly tuned from $0$ to $l \Theta$,
	the soliton will travel across $l n/m$ unit cells ($L=|l np/m|$ sites). 
	To incorporate the effects of a soliton, 
	we define a modified linear Hamiltonian $H^{\text{sc}}(\theta)$ in a supercell consisting of $L$ sites 
	as $H^{\text{lin}}(\theta)$ with an additional on-site potential $\tilde{m}_x(\theta)=g_x|\psi_x(\theta)|^2$~\cite{jurgensenNat2021}.
	We require that $\tilde{m}_x(\theta)=\tilde{m}_{x+L}(\theta)$ so that $H^{\text{sc}}(\theta)$ is
	translationally invariant under a spatial translation of $L$ sites,
	thereby characterized by a momentum $k^{\text{sc}}$. The displacement of the soliton is then
	characterized by the Chern number of a band of $H^{\text{sc}}$ with respect to 
	$k^{\text{sc}}$ and $\tilde{\theta}=\theta^\prime/l$.
	
	For this integer case with $m=1$ and $n=1$, we take $l=4$ so that the supercell contains $L=12$ sites. 
	Figure~\ref{Fig_2}(e) displays the nonlinear on-site potentials $\tilde{m}_x(\theta)$ with $x=1,2,3$, 
	and the others can be obtained based on the relation that 
	$\tilde{m}_{x+3 j}(\theta)=\tilde{m}_{x}(\theta- j \Theta)$ with $j=1,2,3$. 
	We find that the Chern number of the relevant band for the modified linear Hamiltonian 
	corresponding to the soliton in Fig.~\ref{Fig_2}(d) 
	is one, consistent with the displacement of the soliton over one period.
	
	\emph{Nonlinearity-induced fractional Thouless pumping of solitons}.---In contrast to 
	conventional Thouless pumping, nonlinear Thouless pumping can exhibit 
	a distinctive multiperiodic behavior~\cite{jurgensenNP2023}, e.g., a soliton is transported across one unit cell every 
	two periods. This implies that
	the average displacement of a soliton per period is a fraction of a unit cell.
	Previous studies have shown that this fractional displacement is linked to the multi-band structure of solitons,
	with an overall displacement being determined by the sum of the Chern number of these bands in a linear Hamiltonian.
	Consequently, fractional Thouless pumping is precluded if the Chern numbers of all these bands are zero.

	We now show that a soliton can undergo fractional Thouless pumping 
	even when all the bands of the linear Hamiltonian are topologically trivial. 
	For demonstration, 
	we consider the off-diagonal AAH model with $p=5$ and $q=2$. 
	There are five bands in the model, all of which have zero Chern number [see Fig.~\ref{Fig_3}(b)].  
	To induce soliton movement, we vary the hopping strength by modifying $\theta$ 
	as shown in Fig.~\ref{Fig_3}(a).
	Figure~\ref{Fig_3}(c) displays the density distribution of the instantaneous solitons of the instantaneous nonlinear 
	Hamiltonian at each $\theta$, which bifurcate from the lowest two linear bands.
	The figure remarkably illustrates that a soliton is displaced by one unit cell (five sites)
	over {\it two} periods, as also supported by the flow of the soliton's center-of-mass position [see Fig.~\ref{Fig_1}(c)].
	The center-of-mass displacement amounts to exactly $1/2$ unit cell per period,
	a phenomenon we attribute to the parity-time symmetry as detailed in the Supplemental Material. 
	The results are in stark contrast to the multi-band Wannier functions of the linear Hamiltonian,
	which do not exhibit any flow with respect to $\theta$ [Fig.~\ref{Fig_1}(c)].
	
	\begin{figure}
		\includegraphics[width=1\linewidth]{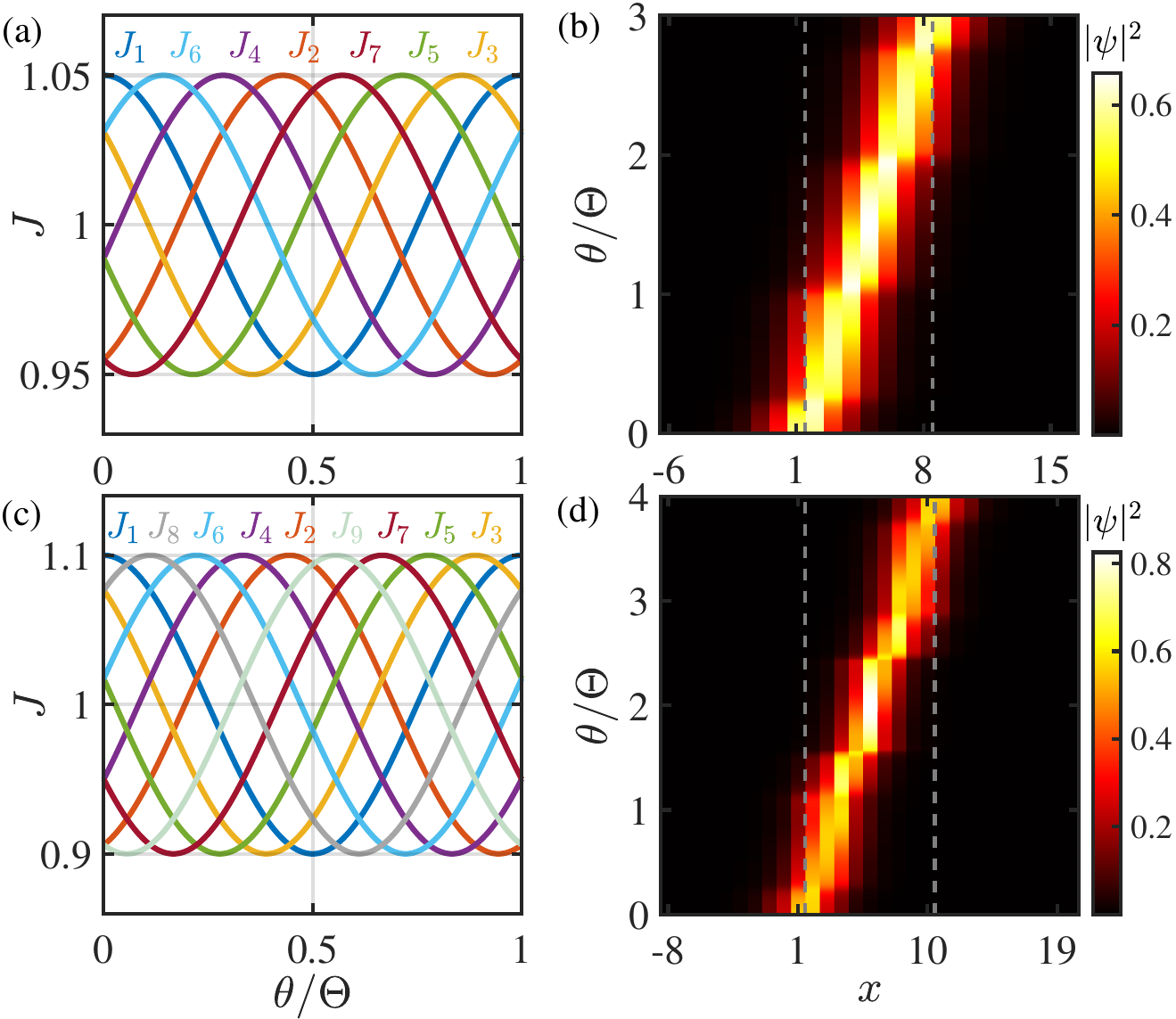}
		\caption{Nonlinearity-induced fractional Thouless pumping where a soliton is pumped across 
			one unit cell (seven sites) over three periods in (a)-(b) and across one unit cell (nine sites) over four
			periods in (c)-(d). 
			[(a) and (c)] Modulation of the hopping strength used to generate the pumping.   
			(b) Three-cycle (d) four-cycle evolution of the density profile $|\psi_x|^2$ of 
			instantaneous solitons that bifurcate from 
			the lowest (b) three or (d) four linear bands. 
			In (a) and (b), we set $p=7$, $q=3$, $N=2.1$, $J_a=0.05$, 
			$\{m_x\}=\{-0.17,-0.17,0.14,-0.13,0.02,-0.13,0.14\}$, 
			and $\{g_x\}=\{-0.77,-0.77,-1.23,-0.94,-1.08,-0.94,-1.23\}$;
			in (c) and (d), we set $p=9$, $q=4$, $J_a=0.1$, $N=2$, 
			$\{m_x\}=\{-0.2,-0.2,0.2,-0.2,0.18,-0.14,0.18,-0.2,0.2\}$, and 
			$\{g_x\}=\{-0.7,-0.7,-1.3,-0.95,-1.3,-1.1,-1.3,-0.95,-1.3\}$.
		}
		\label{Fig_4}
	\end{figure}
	
	Similar to the integer case, we attribute the phenomenon to on-site potentials  
	induced by nonlinear terms. In this case, we have $m=2$ and $n=1$, and 
	set $l=6$ so that a supercell contains $L=15$ sites. The nonlinearly induced on-site potentials 
	$\tilde{m}_x(\theta)$ have a period of $6\Theta$ and are shown in Fig.~\ref{Fig_3}(d)
	[the other $\tilde{m}_x(\theta)$ can be obtained through 
	$\tilde{m}_{x+5 j}(\theta)=\tilde{m}_{x}(\theta-2 j \Theta)$ with $j=1,2$].
	We find that the Chern number of the band associated with the soliton in 
	the modified linear Hamiltonian is $1$, 
	indicating that the Wannier function of this band (an approximation to the soliton solution) 
	is pumped by 15 sites, i.e., 
	three original unit cells after $6\Theta$.
	This is consistent with the displacement of $1/2$ per period.
	
	Nonlinearity-induced fractional Thouless pumping is in fact widely present 
	in the nonlinear off-diagonal AAH model 
	at intermediate strengths of nonlinearity. For example, 
	we also find fractional Thouless pumping of solitons in the case with 
	$(p=7,q=3)$ [Fig.~\ref{Fig_4}(a) and (b)] and $(p=9,q=4)$ [Fig.~\ref{Fig_4}(c) and (d)],
	where all the bands of the linear Hamiltonian have a zero Chern number.
	Pumping is achieved by modulating the hopping strength, as illustrated in Fig.~\ref{Fig_4}(a) and (c).
	The soliton's average displacement per cycle is $1/3$ and $1/4$ unit cell for the former
	and latter, respectively. Equivalently, a soliton is transported across one unit cell (comprising seven or nine sites)
	over three and four cycles, respectively. 
	Similarly, the phenomenon can be explained by the supercell method.
	For the former, we have $m=3$, $n=1$, and set $l=6$ so that  $L=14$;
	for the latter, we have $m=4$, $n=1$, and set $l=8$ so that $L=18$. 
	Similar to the previous cases, the Chern number of the relevant band
	for the modified linear Hamiltonians is found to be $1$.

	The instantaneous soliton solutions we present above are all stable, as confirmed by the stability 
	analysis at each $\theta$. 
	In addition, we have numerically calculated the time evolution of a soliton
	according to Eq.~(\ref{NLSETB}), with a sufficiently slow modulation of $\theta$
	over time. The results are in close agreement with the 
	instantaneous adiabatic predictions. 
	Furthermore, we have utilized the diagonal AAH model to construct the linear Hamiltonian and 
	observed a variety of integer and fractional Thouless pumping induced by nonlinearity 
	(see the Supplemental Material), 
	suggesting the widespread presence of the phenomena.

	In summary, we have theoretically predicted nonlinearity-induced integer 
	and fractional Thouless pumping of solitons in nonlinear 
	off-diagonal and diagonal AAH models.
	For the fractional cases, we show that a soliton can be pumped across $1/2$, $1/3$ and
	$1/4$ unit cell on average over one cycle, despite the fact that the relevant linear energy bands have zero
	Chern numbers. These findings go beyond the previous understanding that
	pumping is determined by the Chern number of the relevant band of the linear Hamiltonian. 
	Given the simplicity of our model---particularly, the modulation of hopping strength which has 
	been experimentally realized~\cite{mukherjeeSci2020,maczewskySci2020,mukherjeePRX2021,jurgensenNat2021,jurgensenNP2023}---we expect that nonlinearity-induced integer 
	and fractional pumping can be experimentally demonstrated. 
	For instance, 
	using a 1D waveguide array with nearest-neighbor evanescent couplings, and applying
	the paraxial approximation, monochromatic light propagation is governed by the 
	discrete nonlinear Schr{\"o}dinger Eq.~(\ref{NLSETB}). In this context, 
	$\psi_x$ describes the electrical field envelope at site $x$, and
	$t$ represents the propagation distance. The parameter $\theta$ is slowly 
	modulated along the distance. 
	To realize our model, one can choose 
	waveguides with different shapes and materials to obtain site-dependent propagation constants and 
	nonlinear refractive index coefficients, which correspond to on-site potentials $m_x$ and 
	nonlinear coefficients $g_x$, respectively. In addition, modulating the spacing between adjacent 
	waveguides in $t$ allows for $\theta$-dependent off-diagonal nearest-neighbor hoppings. 
	In our numerical simulations, we find that varying $\theta$ according to $\theta=t\Theta/T$ 
	with $T=(2,8,4,5)\times 10^{3}$ for the $(p,q)=(3,1), (5,2), (7,3), (9,4)$ cases 
	can ensure good adiabaticity. 
	In experiments, using a typical parameter $J_0=0.15 \text{ mm}^{-1}$~\cite{jurgensenNat2021,jurgensenNP2023}, 
	we estimate propagation distances of $13.3$ m, $53.4$ m, $26.7$ m, and $33.3$ m 
	per period for these cases, respectively.

	\begin{acknowledgments}
		We thank J.-X. Wen for helpful discussions.
		This work is supported by the National Natural Science Foundation of China (Grant No. 12474265, 11974201,
		12374247, and 11974235)
		and Innovation Program for Quantum Science and Technology (Grant No. 2021ZD0301604).
		Y. Zhang is also supported by the Shanghai
		Municipal Science and Technology Major Project (Grant
		No. 2019SHZDZX01-ZX04).
		We also acknowledge the support by center of high performance computing, Tsinghua University.
	\end{acknowledgments}   
	
	%
	
	\begin{widetext}
			\setcounter{equation}{0} \setcounter{figure}{0} \setcounter{table}{0} %
		\renewcommand{\theequation}{S\arabic{equation}} \renewcommand{\thefigure}{S%
			\arabic{figure}}
		
		In the Supplemental Material, we will prove that the parity-time symmetry can protect the 
		half nonlinear Thouless pumping
		in Section S-1,
		and provide the results for
		nonlinearity-induced Thouless pumping of solitons in the diagonal AAH model
		in Section S-2.
		
		\section{S-1. Half nonlinear Thouless pumping protected by the parity-time symmetry}
		For fractional nonlinear Thouless pumping, a soliton returns to its initial shape with a 
		displacement of $n$ unit cells after a minimum of $m$ periods. Thus,
		an average displacement per period is $n/m$ unit cells. However, this does not mean 
		that the displacement over one cycle (i.e., $\theta=\Theta$) is exactly $n/m$ unit cells. 
		In fact, it usually differs from this exact value. Interestingly, we find that the 
		displacement at $\theta=m\Theta/2$ is exactly $n/2$ unit cells as shown in Fig. 1(c). In this section,
		we will prove that this fact is protected by the parity-time symmetry. Here, the parity operation refers to
		the reflection one. Consider the reflection center at $x=x_r$ (in our case $x_r=1.5$) and a unit cell containing $p$ sites.
		With the party-time symmetry in 1D, we require that
		\begin{align}
			\label{PT}
			[H^{\textrm{lin}}(-\theta)]_{x,x^\prime}&=[H^{\textrm{lin}}(\theta)]_{2x_r-x,2x_r-x^\prime}\\ \nonumber
			g_{x}&=g_{2x_r-x}.
		\end{align}
		Since the system is also translationally invariant in space and periodic with respect to $\theta$, we write
		the above relations as
		\begin{align}
			\label{PT}
			[H^{\textrm{lin}}(l \Theta-\theta)]_{x,x^\prime}&=[H^{\textrm{lin}}(\theta)]_{j p+2x_r-x,j p+2x_r-x^\prime}\\ \nonumber
			g_{x}&=g_{j p+2x_r-x}
		\end{align}
		for any integers $l$ and $j$. Let $\psi_{x}(\theta)$ be a soliton solution to the nonlinear Hamiltonian at $\theta$ so that
		\begin{align}
			\label{soliton}
			\sum_{x^\prime}[H^{\textrm{lin}}(\theta)]_{x,x^\prime}\psi_{x^\prime}(\theta)+g_{x}|\psi_{x}(\theta)|^2\psi_{x}(\theta)=\mu\psi_{x}(\theta),
		\end{align}
		where $\mu$ is a nonlinear eigenvalue. By changing the spatial index $x$ to $j p+2x_r-x$ in Eq.~(\ref{soliton}), we derive that
		\begin{align}
			\label{solitonPT}
			\mu\psi_{j p+2x_r-x}(\theta)&=\sum_{x^\prime}[H^{\textrm{lin}}(\theta)]_{j p+2x_r-x,x^\prime}\psi_{x^\prime}(\theta)+g_{j p+2x_r-x}|\psi_{j p+2x_r-x}(\theta)|^2\psi_{j p+2x_r-x}(\theta)\\ \nonumber
			&=\sum_{x^\prime}[H^{\textrm{lin}}(l \Theta-\theta)]_{x,j p+2x_r-x^\prime}\psi_{x^\prime}(\theta)+g_{x}|\psi_{j p+2x_r-x}(\theta)|^2\psi_{j p+2x_r-x}(\theta)\\ \nonumber
			&=\sum_{x^\prime}[H^{\textrm{lin}}(l \Theta-\theta)]_{x,x^\prime}\psi_{j p+2x_r-x^\prime}(\theta)+g_{x}|\psi_{j p+2x_r-x}(\theta)|^2\psi_{j p+2x_r-x}(\theta),
		\end{align}
		where we have used the relations in Eq.~(\ref{PT}). The equation tells us that
		$\psi_{j p+2x_r-x}(\theta)$ is an instantaneous soliton solution at $\theta^\prime =l \Theta-\theta$.
		The soliton's center-of-mass position is given by
		\begin{align}
			\label{xcanother}
			x_c(l \Theta-\theta)&=\left[ \sum_{x}x|\psi_{j p+2x_r-x}(\theta)|^2 \right]/N \\ \nonumber
			&=\left[ \sum_{x}(j p+2x_r-x)|\psi_{x}(\theta)|^2 \right]/N \\ \nonumber
			&=j p+2x_r-x_c(\theta).
		\end{align}
		We therefore arrive at
		\begin{equation}
			x_c(l \Theta-\theta)+x_c(\theta)=j p+2x_r.
		\end{equation}
		For an evolution which connects a soliton at $\theta=0$ and one at $\theta=l\Theta$, we have
		\begin{align}
			\label{xcconst}
			x_c(l \Theta-\theta)+x_c(\theta)=x_c(l \Theta)+x_c(0).
		\end{align} 
		
		Given that a soliton $\psi_{x}(0)$ evolves to $\psi_{x}(m\Theta)=\psi_{x-np}(0)$ after $m$ periods, 
		we have $x_c(m\Theta)=x_c(0)+np$. Based on Eq.~(\ref{xcconst}), by setting $l=m$, we obtain
		\begin{align}
			\label{halfpumping_1}
			x_c(m\Theta-\theta)+x_c(\theta)=2x_c(0)+np.
		\end{align}
		We thus derive that the center-of-mass position at $\theta=m\Theta/2$ is $x_c(m\Theta/2)=x_c(0)+np/2$,
		indicating that the displacement in units of the length of a unit cell at $\theta=m\Theta/2$ is
		\begin{align}
			\label{halfpumping_2}
			\delta x_c(m\Theta/2)=\frac{1}{p}[x_c(m\Theta/2)-x_c(0)]=n/2.
		\end{align}
		
		\section{S-2. Nonlinearity-induced Thouless pumping of solitons in the diagonal AAH model}
		In this section, we will show that nonlinearity can also induce Thouless pumping of solitons in 
		a nonlinear diagonal AAH model. The model can still be described by Eq. (1) in the main text.
		Unlike the off-diagonal AAH model, the hopping strength is independent of the spatial index
		and $\theta$; in fact, we set $J=1$ as the unit of energy. Instead,
		on-site potentials $m_x(\theta)$ are periodically modulated in $x$ and 
		$\theta$,
		taking the following form
		\begin{equation}
			\label{mx}
			m_{x}(\theta)=\delta m_x-m_a\cos\left(2\pi \theta/\Theta-2\pi(x-1)q/p\right),
		\end{equation} 
		where $m_a$ is the oscillation amplitude of modulated potentials, and $\delta m_x$ is the central potential at site $x$. 
		Here, $q$ and $p$ are positive integers with $q$ being prime to $p$ and $q<p/2$. 
		Regarding nonlinearity, as in the main text, we choose site-dependent nonlinear coefficients $g_x$ with a 
		period of $p$ sites.
		
		By modulating on-site potentials according to Eq.~(\ref{mx}), we observe both integer and fraction Thouless
		pumping for a soliton induced by nonlinearity. The average displacement per cycle for these pumping is 
		$-2$, $1/2$, $1/3$ and $1/4$ unit cells, respectively, as shown in Fig.~\ref{Fig_AAH}. 
		The solitons for all the cases bifurcate from topologically trivial linear bands with zero Chern numbers.
		
		\begin{figure}[htp]
			\includegraphics[width=1\linewidth]{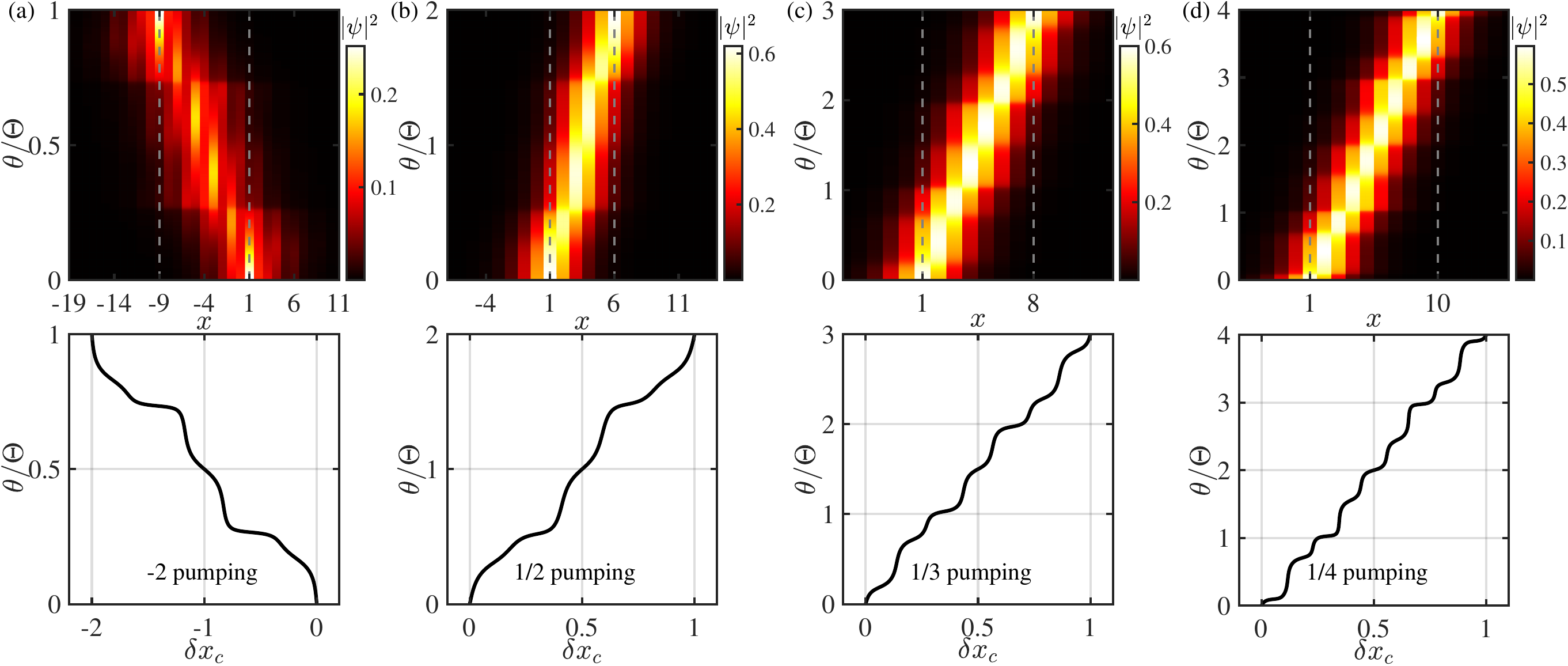}
			\caption{Nonlinearity-induced Thouless pumping in the diagonal AAH model. 
				Upper panels: The evolution of the density distribution $|\psi_x|^2$ of instantaneous 
				solitons bifurcating from topologically trivial linear bands with zero Chern numbers. 
				The dashed lines indicate the initial and final center-of-mass positions of a soliton,
				corresponding to its original location and its displacement by one unit cell, respectively. 
				Lower panels: The trajectory of relative center-of-mass positions of the corresponding 
				solitons in the upper panels.	
				A soliton is pumped across $-2$ unit cells in (a) and $1$ unit cell in (b)-(d)  
				over one period in (a), two periods in (b), three periods in (c), and four periods in (d).
				In (a), $p=5$, $q=2$, $m_a=0.5$, $N=0.9$, $\{\delta m_x\}=\{-0.1,0,0,0,0\}$, and 
				$\{g_x\}=\{-1,-1,-1,-1,-1\}$.
				In (b), $p=5$, $q=2$, $m_a=0.05$, 
				$N=2$, $\{\delta m_x\}=\{-0.05,0.05,0,0,0.05\}$, and $\{g_x\}=\{-0.95,-1,-1,-1,-1\}$.
				In (c), $p=7$, $q=3$, $m_a=0.05$, $N=2$, $\{\delta m_x\}=\{-0.05,0.05,-0.03,0.03,0.03,-0.03,0.05\}$, 
				and $\{g_x\}=\{-0.92,-1.05,-0.95,-1.05,-1.05,-0.95,-1.05\}$. 
				In (d), $p=9$, $q=4$, $m_a=0.05$, $N=2$, $\{\delta m_x\}=\{-0.05,0.05,-0.03,0.03,-0.02,-0.02,0.03,-0.03,0.05\}$, and $\{g_x\}=\{-0.92,-1.08,-0.96,-1.04,-0.96,-0.96,-1.04,-0.96,-1.08\}$.}
			\label{Fig_AAH}
		\end{figure}
	
	\end{widetext}
	
\end{document}